\documentclass[10pt,twocolumn,prb]{revtex4}

\usepackage{bm,graphicx,amsfonts,amssymb}

\begin{document}

\title{A constructive approach to the special theory of relativity}
\author{D. J. Miller}
\email[Email address: ]{davmille@arts.usyd.edu.au}
\affiliation{Centre for Time, University of Sydney, Sydney NSW 2006, Australia and \\ School of Physics, University of New South Wales, Sydney NSW 2052}

\date{\today}

\begin{abstract}
Simple physical models of a measuring rod and of a clock are used to demonstrate the contraction of objects and clock retardation in special relativity. It is argued that the models could help in promoting student understanding of special relativity and in distinguishing between dynamical and purely perspectival effects.
\end{abstract}

\pacs{03.30.+p, 01.70.+w}
\maketitle


\section{Introduction}

Several factors make the special theory of relativity (STR) difficult to understand on first acquaintance. One factor is the strongly counterintuitive nature of some of the important results, especially the relativity of simultaneity, time dilation and length contraction. Another factor is that STR is almost invariably taught, both in the classroom and textbooks, in a different way from all the other areas of physics in the undergraduate curriculum except for classical thermodynamics. This is because the results of STR are obtained by deduction from the relativity postulate and the light postulate (either in the forms originally proposed, or versions of them). That is, STR is usually taught as a so-called ``principle" theory. Teaching STR that way is especially problematic because, unlike the case of classical thermodynamics which is also taught as a principle theory, the two postulates or principles in the case of STR are strongly counterintuitive when taken together. That is because the natural models of light as either an electromagnetic wave propagating in a medium (the ether) or a photon emitted by its source are both apparently ruled out by the relativity postulate and the light postulate respectively. Therefore it is particularly unsatisfactory that it is the peculiar property of the speed of light (often as it bounces about in stationary or moving railway carriages) which is relied on in many introductory treatments of STR to introduce the student to time dilation, the relativity of simultaneity and length contraction.

In other areas of undergraduate physics, results are deduced on the basis of dynamical laws involving specific physical forces or fields. That is, the other areas are taught as ``constructive" theories. As a consequence, a part of the difficulty of understanding STR are unanswered questions at the back of one's mind like: surely the contraction of this measuring rod is the result of a force? what changes inside the clock to make it tick slower when I see it moving? 

Minds differ on whether STR is best regarded as a principle or a constructive theory.  Harvey Brown's recent book \cite{brownbook} argues in favour of the constructive approach. Recent arguments in favour of a principle approach can be found in the work of Janssen \cite{janssen} and of Norton \cite{norton}. Even if one feels that it is best to approach STR as a principle theory and derive the results of STR from principles by kinematic arguments, it is common ground in most of the literature that a constructive account of the phenomena of STR based on dynamical arguments can also be given. Unfortunately, this point is not made clear in many textbooks and, even in some of the very best textbooks, there are comments which might inadvertently imply the contrary to some readers. Generally the constructive approach, based as it is on the dynamics of the problem at hand, provides a causal explanation and a better physical insight into the problem. \cite{brownbook, bell} Thus there is no impediment against introducing the concepts of STR via a constructive approach even if one is going to argue in hindsight that the concepts are best seen to flow directly from the principles.

In a well-known article, Bell \cite{bell} advocated the constructive approach to teaching relativity and analyzed a physical model that could be used to illustrate the method. A problem with Bell's model is that it cannot be solved analytically and so the impact of the result is obscured somewhat by the need for a numerical solution. Another problem with it is that the relativistic form of Newton's second law needed to be anticipated at the outset.

The main aim of the present work is to present a more convenient example and to sketch very briefly how the main results of STR can be obtained from a constructive approach to the subject. Some remarks on the constructive/principle approaches to STR are provided in Sec.~III.

\section{A constructive approach}

The development of most areas of physics begins with the discovery or proposal of the relevant laws of physics and proceeds by the elaboration of their relevant consequences. A constructive approach to STR should proceed in the same way, i.e. with a physical law that can be used to investigate the behaviour of measuring rods and clocks. The obvious candidate now, as it was at the time of Fitzgerald, Lorentz and others who also took a constructive approach to some of the problems dealt with by STR, is Maxwell's theory of electromagnetism. It appears \cite{einsteinmaxwell} that one of the reasons that Einstein preferred a principle approach to STR was that he was concerned that Maxwell's theory might need to be modified in the light of contemporary developments. More than a century later, we need not share those qualms. 

In order to make the constructive derivation of length contraction and clock retardation as transparent as possible, it is desirable to construct a measuring rod and a clock from a very simple system such as a system of point charges in equilibrium. A problem is that there are no stable configurations of charges that can be maintained by the interaction of the charges with the electric and magnetic fields they produce. Diamagnetic materials are an exception to this rule \cite{diamag}, which is an extension of Earnshaw's theorem \cite{earnshaw}, but invoking diamagnetism is undesirable because of the need for simplicity. As emphasized by Swann \cite{swann} in the present context, material objects are in equilibrium due to quantum mechanical effects but considering those is also ruled out by the need for simplicity. Fortunately there are simple qualitative arguments \cite{taylor} that show that lengths of material objects transverse to the direction of their motion are independent of their speed, i.e. transverse length is the same for all inertial observers.  We can rely on that to stabilize a system of charges which can then be used to derive length contraction and clock retardation.

Our simple ``measuring rod" is shown in Fig.~1(a). It is the distance between the $+Q$ charges which are repelled by each other and attracted by two charges $-q$ attached to the ends of a non-conducting bar of length $2L$ with relative permittivity 1. We only consider motion along the $x$-axis which is perpendicular to the length $2L$ so, on the basis of the qualitative arguments mentioned above, all observers agree on the separation $2L$ (and do so independently of the definition of simultaneity which is therefore not required at this stage). The two charges $+Q$ are in stable equilibrium for displacements along the $x$-axis but unstable equilibrium for small displacements perpendicular to the $x$-axis so it best to imagine that the $+Q$ charges are inside a transparent and frictionless tube lying along the $x$-axis.

\begin{figure}[b]
\begin{center}
\includegraphics[scale=0.4]{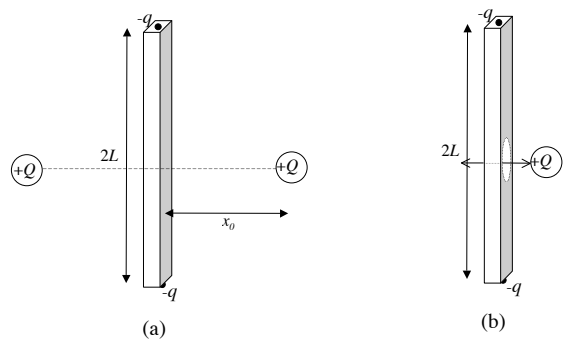}
\caption{(a) The distance between the $+Q$ charges in their equilibrium positions represents the measuring rod. The $-q$ charges are held in position by an insulating rod of length $2L$. When $Q=2\sqrt{2}q$ and the charges are at rest with respect to the observer the separation of the $+Q$ charges is also $2L$. When the whole arrangement is set in motion in the horizontal direction, the distance between the $+Q$ charges is contracted. (b) The  periodic motion of the $+Q$ charge after it slightly displaced from the center of the rod represents a clock. When the whole arrangement is set in motion in the horizontal direction, the period of the motion becomes greater so the ``clock" becomes retarded.}
\end{center}
\end{figure}

Perhaps it s worth noting that in order to write down Maxwell's equations and perform the calculations, we need a system of coordinates which, in turn, requires rods and synchronised clocks to set up. We therefore must first assume that rods and clocks exist and that the clocks can be synchronised, say by slow clock transport, and then show that the existence of such rods and clocks is consistent with the theory. 

\subsection{Contraction of a moving measuring rod}

If the charges are at rest in an inertial frame $S$, only electrostatic forces are involved. In equilibrium, the $+Q$ charges occupy the positions where the electric field is zero and, by symmetry, these are equidistant from the rod at $\pm x_0$ along the $x$-axis which bisects the rod. Also by symmetry, only the $x$-component of the electric field on the $x$-axis $E_x\left (x_0 \right)$ can be non-zero. Thus $x_0$ can be found from the condition 
\begin{equation}
E_x\left (x_0\right)=\frac{1}{4\pi\epsilon_o}\left [\frac{Q}{4x_0^2}- \frac{2qx_0}{\left (x_0^2 + L^2\right)^{3/2}}\right ]=0
\end{equation}
where $\epsilon_o$ is the permittivity of free space. If $Q=2\sqrt{2}q$, Eq.~(1) is satisfied when $x_0=L$ and so the equilibrium separation of the $+Q$ charges is $2L$ in the rest frame of the charges.

To consider what happens when the charge assembly in Fig.~1(a) moves speed $v$ in the $x$-direction with respect to $S$, we require the electric and magnetic fields of charges in uniform motion. While some students in a course on STR may not have met the required expressions, there are relatively simple derivations of them. \cite{dmit,hu} Of course, one must not assume at this stage any of the results of STR so it is essential to the present approach that the required expressions are obtained directly from Maxwell's equations. Alternatively students could be asked to accept the expression for the electric field of a moving charge as given. Intuitively it is not unreasonable that the electric field lines become ``squashed" in the direction of movement of a charge (and the magnetic field lines can be derived from those using $\bm{B}=\bm{v} \times \bm{E}/c^2$ where $\bm{v}$ is the velocity of the source of $\bm{E}$ and $c$ is the speed of light).

When all four charges move with speed $v$ in the $x$-direction, the electric and magnetic fields at a vector distance $\bm{r}$ from the current (not retarded) position at time $t$ of a charge $q$ moving with velocity $\bm{v}=v\bm{\hat{x}}$ are \cite{griffeb}
\begin{eqnarray}
\bm{E}\left (\bm{r},t \right ) & = & \frac{q}{4\pi\epsilon_o}\frac{\hat{\bm{r}}}{r^2}\frac{1-\beta^2}{\left (1-\beta^2\sin^2\theta\right )^{3/2}} \\
\bm{B}\left (\bm{r},t \right ) & = & \frac{1}{c^2}\left (\bm{v} \times \bm{E}\left (\bm{r}, t \right )\right ) 
\end{eqnarray}
where $\beta=v/c$ and $\theta$ is the angle between $\bm{r}$ and the $x$-axis.
Once again the equilibrium separation can be determined from the condition that the fields at each of the $+Q$ charges are zero. There is no nett magnetic field at the $+Q$ charges and the separate conditions for each of the $+Q$ charges show that they are still equidistant from the rod by a distance we can call $x_v$. The equilibrium separation $2x_v$ of the moving $+Q$ charges can be found from  
\begin{eqnarray}
\lefteqn{E_x\left (x_v\right )  =  \frac{\left (1-\beta^2\right )}{4\pi\epsilon_o} } \nonumber \\
& & \left [\frac{Q}{4x_v^2}- 
\frac{2qx_v}{\left (x_v^2 + L^2\right )^{3/2}\left (1-\beta^2\frac{L^2}{x_v^2 + L^2}\right )^{3/2}}\right ]=0
\end{eqnarray}
which is satisfied when
\begin{equation}
x_v=\sqrt{1-\beta^2}L=L/\gamma
\end{equation}
where $\gamma=1/\sqrt{1-\beta^2}$. Hence, when the charge assembly is moving with constant speed $v$, the moving ``measuring rod" represented by the equilibrium separation of the $+Q$ charges is reduced to $2\sqrt{1-\beta^2}L$ from their separation, the length $2L$ of the ``measuring rod", when at rest.

\subsection{Lorentz transformation}

As shown in Fig.~1(b), the above arrangement can be modified to produce a clock by removing one of the $+Q$ charges and moving the other $+Q$ charge between the $-q$ charges (the vertical bar needs to have a suitable gap drilled through it to accommodate the $+Q$ charge). If the centre charge is given a horizontal displacement from the point midway between the $-q$ charges, it will experience a restoring force and execute a periodic motion which can be used to measure time. In order to calculate the period, it is necessary to have the correct form of Newton's second law, i.e. the relativistic form of the second law. One can reach that result in a logical manner in the present approach by first noting that length contraction is inconsistent with the Galilean transformation of co-ordinates between inertial frames. That motivates a re-consideration of the transformation of co-ordinates leading to the correct (Lorentz) transformation. The correct (relativistic) form of the laws of mechanics then follow as in any course on STR.

\begin{figure}[b]
\begin{center}
\includegraphics[scale=0.4]{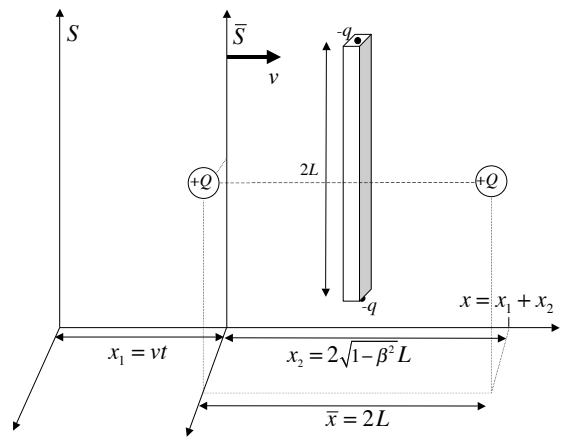}
\caption{The measuring rod apparatus is at rest in frame $\overline{S}$ which moves horizontally to the right with speed $v$ with respect to frame $S$. At time $t$ in $S$, the trailing edge of the ``rod" (left $+Q$ charge) is at the origin of $S$ and $\overline{S}$ and the leading edge (right $+Q$ charge) is at $x$ in $S$ and at $\overline{x}$ in $\overline{S}$.}
\end{center}
\end{figure}

A standard text-book argument \cite{griffiths} can be adapted to the present case. Let the measuring rod be at rest in frame $\overline{S}$ with one end at the origin and the other at $\overline{x}=2L$ as shown in Fig.~2. At $t=0$ in $S$, the origins of $S$ and $\overline{S}$ coincide with one end of the rod, and the other end of the rod is located at $x$ in $S$ and $\overline{x}$ in $\overline{S}$ at the same time. Note again this requires clock synchronisation in only a single frame which can be carried out by a variety of methods, e.g. slow clock transport, and which does not anticipate the relativistic results involving time which we are in the process of deriving. In terms of the units of length used in $S$, the rod is contracted to $2\sqrt{1-\beta^2}L$ using Eq.~(5) while in $\overline{S}$ the rod is at rest and so $\overline{x}=2L$ from Eq.~(1). From inspection of Fig.~2, the position of the $+Q$ charge on the right is $x=vt+2\sqrt{1-\beta^2}L$ and $\overline{x}=2L$, which leads to
\begin{equation}
\overline{x} = \gamma \left (x-vt\right ). \label{xbar}
\end{equation}

If we repeat a similar argument from the point of view of reference frame $\overline{S}$, we find \cite{reciprocity}
\begin{equation}
x = \gamma \left (\overline{x}+v\overline{t}\right ).
\end{equation}
Substitution of $\overline{x}$ from Eq.~(6) into Eq.~(7), leads directly to
\begin{equation}
\overline{t} = \gamma\left (t-\frac{v}{c^2} x\right ). \label{tbar}
\end{equation}
Together with the general argument about the co-ordinates transverse to the motion remaining unchanged, Eqs.~(\ref{xbar}) and (\ref{tbar}) constitute the familiar Lorentz transformation (LT).

Once the LT has been obtained in the constructive approach to STR, the rest of STR follows, including, relevantly for us below, the relativistic form of Newton's second law:   
\begin{equation}
\bm{F}\left (\bm{x}\right ) = \frac{\partial \bm{p}}{\partial t}= \gamma^{3} m  \frac{d^2\bm{x}}{dt^2}.
\end{equation}
 
\subsection{Slowing of a moving clock}

Although clock retardation follows from the LT alone, the constructive approach requires us to show that a material clock can be constructed within the theory and that it behaves in the expected manner.

When the middle charge $+Q$ in Fig.~1(b) is displaced from its equilibrium position by $\bm{x}_v$, the relativistic restoring force $\bm{F}\left (x_v \right)$, with the charges moving at speed $v$ with respect to $S$, is 
\begin{eqnarray}
\bm{F}\left (x_v \right ) & = & - \frac{Q\left (1-\beta^2\right )}{4\pi\epsilon_o} \nonumber \\ 
 & & \frac{2qx_v}{\left (x_v^2 + L^2 \right )^{3/2}
\left (1-\beta^2\frac{L^2}{x_v^2 + L^2}\right )^{3/2}} \hat{\bm{x}}_v \\
 & \sim & - \gamma \frac{qQ}{4\pi\epsilon_oL^3} \bm{x}_v \;\;\;\; \text{for} \;\;\;\; x_v<<L.
\end{eqnarray}
Using the relativistic form of Newton's second law in Eq.~(9), the centre charge executes simple harmonic motion under the influence of the restoring force with a period
\begin{equation}
T_v=2 \pi \gamma \sqrt{\frac{4\pi\epsilon_0L^3m}{qQ}}=\gamma T_0
\end{equation}
where $T_0=2\pi \sqrt{4\pi\epsilon_0L^3m/qQ}$ is the period when the clock is at rest. This motion constitutes a simple clock. It runs more slowly by the factor $\sqrt{1-\beta^2}$ when it is moving in $S$ than when it is at rest and it is reasonable to suggest that more elaborate moving clocks will run slowly for similar reasons.

\section{Discussion}

It would be possible to begin a course of STR with the above dynamical derivations of length contraction, time dilation and the LT. Having reached that point, the course would follow the more traditional development although it would be more consistent to derive results using dynamical arguments wherever possible.

An advantage of the arrangement of charges in Fig.~1 is that they could have macroscopic dimensions and could be imagined to be on a laboratory bench top (the arrangement in Fig.~1 is reminiscent of the twin ``pith" balls which serve as an elementary demonstration of charge repulsion). Another significant advantage is that they don't involve expressly the invariance of the speed of light for all inertial observers which is one of the counterintuitive consequences of STR \cite{postulates} and therefore should be avoided in demonstrating other counterintuitive consequences of STR. Perhaps the main advantage is that the length contraction of an object that changes speed is derived within a single frame and does not involve the comparison of the synchronization of clocks in two different inertial frames. This eliminates the notion that the length contraction that occurs when a physical object changes frames somehow involves the relativity of simultaneity.

The rod and clock models are reasonable because they are analogous to the way one imagines the ion cores are held in position in real materials. Of course, treatment of real materials considered as an assembly of ion cores and electrons requires a quantum mechanical solution. On the other hand, the Hamiltonian involves only the electromagnetic field used in the proposed models. It recent times it has become possible to calculate the lattice constant of simple solids from genuinely \textit{ab initio} calculations. \cite{zunger} There does not appear to be such a calculation for a solid moving at relativistic speeds. In any case, the necessary calculations are clearly too complicated to be included in an undergraduate course in STR.

An alternative approach is to consider the measuring rod as an elastic solid. Davidon \cite{davidon} has dealt thoroughly with the kinematics and dynamics of moving and accelerated rods from that point of view and derived many interesting results, including length contraction, by considering the dynamics in a single frame. Unfortunately, the analysis is not simple enough to be suitable for an introduction to STR.

\subsection{Dewan-Beran thought experiment}

It is interesting to apply the thought experiment originally proposed by Dewan and Beran \cite{dewan} (also referred to in Bell's article \cite{bell}) to the configuration of charges in Fig.~1(a). Firstly remove the $+Q$ charges from the spheres when they are in their equilibrium configuration at rest in $S$. The spheres remain in position. As measured in $S$, let all the components of the apparatus be simultaneously subject to a uniform acceleration \cite{gravfield} in the $+x$-direction for the same period so that their velocity changes from zero to $v$ and they end up at rest in $\overline{S}$. When they reach rest in $\overline{S}$, both the spheres remain separated by $2L$ as measured in $S$. \cite{dewan} Now replace the $+Q$ charges. The re-charged spheres will move to their new equilibrium position separated by $2\sqrt{1-\beta^2}L$ as measured by $S$.

All inertial observers \cite{xdirection} agree on some aspects of this sequence of events and disagree on other aspects. All the inertial observers agree that all parts of the apparatus started at rest in $S$ and eventually ended at rest in $\overline{S}$ and that all parts of the apparatus were given the same acceleration for the same time. All inertial observers agree that the pair of spheres move closer together when re-charged and that the ratio of the initial (uncharged) to the final (charged) separations of the spheres in $\overline{S}$ is $\gamma$. Inertial observers disagree on the magnitudes of various quantities: the velocities of $S$ and $\overline{S}$, the acceleration, the period of acceleration and the lengths separating the charged and the uncharged spheres in $S$ and $\overline{S}$. They also disagree on the timing of the beginning of the period of acceleration of the various parts of the apparatus. For example, from the point of view of $\overline{S}$, who initially observes the apparatus receding at speed $-v$, the trailing uncharged sphere is judged to begin accelerating (in the opposite direction to the velocity) before the leading uncharged sphere. As a consequence the separation between the charges increases so that when both charges reach rest in $\overline{S}$, they are separated by $2\gamma L$ according to $\overline{S}$. When the $+Q$ charges are replaced, $\overline{S}$ observes them contracting to a separation of $2L$.

This thought experiment separates in time, and therefore makes explicit, the two dynamical processes that are in operation when a physical object is transferred between inertial frames: (i) the external force producing the required acceleration and (ii) the internal forces which maintain the object in equilibrium. Feinberg has considered this question in a more general way. \cite{feinberg} Of course, in a normal object in which the charges are left in place, the internal forces act continuously so as to maintain, or attempt to maintain, the charges in their instantaneous equilibrium positions. The effect of the internal forces depends on the mode of acceleration (the timing of the start of the acceleration, whether the acceleration is constant, etc) involved in changing the velocity of the object.

The direction of the acceleration in relation to the direction of the velocity makes a qualitative difference. We have just dealt with the case of the acceleration in the opposite direction to the velocity, which is always what is required from the point of view of the receiving frame. Some observers will judge that the assembly with uncharged spheres has to speed up to reach $\overline{S}$ from $S$, i.e. the acceleration is in the same direction as the velocity of $S$. For those observers the trailing end of the apparatus begins, as before, to accelerate first but now it is catching up with the other sphere so that the separation of uncharged spheres decreases over its original value in $S$. Despite their disagreement on the magnitude of the separation of the uncharged spheres and whether that separation increased or decreased as the result of the change of frames, as already mentioned, all observers agree that the spheres move together when the charges are replaced and that they move together by the same factor compared with their separation when uncharged. This demonstrates that the fractional change in length with change in velocity caused by the forces that maintain the equilibrium configuration are real in the sense that all inertial observers agree on them.

For completeness, consider the equal acceleration of all parts of the apparatus originally at rest in $S$ is begun simultaneously from the point of view of $\overline{S}$, the receiving frame. The uncharged spheres reach $\overline{S}$ with their initial separation observed by $\overline{S}$, namely $2\sqrt{1-\beta^2}L$. On re-charging in $\overline{S}$, the $+Q$ charges move further apart by a factor $\gamma$ to reach the points of equilibrium separated by $2L$. All inertial observers agree on this movement (while disagreeing, as before, about the magnitudes of the separations involved). Thus if a rod is transferred between frames, the movement of the charges in the rod from their old equilibrium separations in the original frame to their new equilibrium separations in the final frame, under the influence of the dynamical forces within the rod, may produce an expansion or a contraction depending on the manner in which the rod is accelerated between frames. It is even possible to contrive the acceleration in such a way that the uncharged spheres in our example (and the ion cores in a rod in the real world) reach $\overline{S}$ in their equilibrium positions in that inertial frame. Then re-charging the spheres would cause no change in their separation at all. Taylor and Wheeler describe such an acceleration of a measuring rod between frames.\cite{taylor2} 

\subsection{Dynamical v perspectival effects}

The treatment of the thought experiment in Sec.~IIIA relies on the constructive or dynamical approach to STR. It exposes clearly several aspects of STR which remain implicit or hidden in a purely principle or kinematic approach to STR. Firstly it distinguishes between perspectival and dynamical effects. The changes in the separations of the spheres in their destination inertial frame when they are uncharged and then re-charged, show that when a physical object in equilibrium changes frames, there are physical changes in it which can be calculated from a consideration of the forces which keep it in equilibrium and it is those changes which lead to the satisfaction of the Lorentz transformation conditions.

On the other hand when an observer changes frames (or when we compare the results of observers in different frames), there are no dynamical effects in the physical object being observed---the differences among the observers are due to their different perspectives. In the above thought experiment, the disagreements among the inertial observers about the magnitudes of quantities and the relative timing of the beginning and ends of the acceleration period for the various parts of the apparatus are perspectival. They have no dynamical explanation, nor do they require one. This point can be confusing because the factors involved in the perspectival and dynamical effects involve the same numerical values (determined by $\gamma$). When the measuring rods and clocks are moved between inertial observers they suffer dynamical changes. When each observer use their dynamically altered rods and clocks to make measurements, it is not surprising that their results differ and that they differ by the same factors that are involved in the dynamical changes.

While the point might appear trite and is obvious from the constructive approach to STR, failure to appreciate the differences between moving a physical object between frames and moving an observer between frames sometimes produces confusion. For example, it has been used recently as argument against the constructive approach to STR. \cite{martinez} The difference between transferring physical objects between inertial frames and transferring observers between inertial frames is summarised in Table I where the length of a physical object in equilibrium is labeled a ``connected" length to distinguish it from the distance between two objects which are independent of each other which is an ``unconnected" length. One sees that for the perspectival changes resulting from a change of inertial frame by an observer (or the comparison of observations of two observers), a distance appears to change by $\gamma$ whether or not the distance is a connected or an unconnected length in the above sense. However when physical objects are transferred between inertial frames, the connected lengths change in the ratio $\gamma$ but the unconnected lengths change by a factor which depends on the mode of acceleration involved in transferring the objects between inertial frames. The last point has been illustrated by the measuring rod apparatus; e.g. in Sec.~IIIA, the separation of the uncharged spheres is an unconnected length and on transfer from $S$ to $\overline{S}$, their separation remains at $2L$ but when charged spheres are transferred in an identical manner, and remained in equilibrium, their final separation is $2\sqrt{1-\beta^2}L$.

\begin{table}
\caption{The ratio of lengths involving two different inertial frames moving at relative speed $v$ along the direction of the lengths with $\gamma = \left (1-v^2/c^2 \right )^{-\frac{1}{2}}$. The ``connected" lengths refer to lengths of objects in equilibrium and ``unconnected" lengths refer to distances between independent objects. The first row is a perspectival effect: the ratio of the length observed from one inertial frame to the length observed from the other inertial frame by two different observers. The second row is due to dynamical effects. The ratio of the unconnected lengths is variable because it depends on the mode of acceleration of the physical objects between the inertial frames.}
\begin{ruledtabular}
\begin{tabular}{lcc}
Ratio of lengths when & Connected & Unconnected \\
& lengths & lengths \\
\hline
Observer changes frames & $\gamma$ & $\gamma$ \\
Objects change frames & $\gamma$ & Variable\\
\end{tabular}
\end{ruledtabular}
\end{table}

The LT can be considered either as an active or a passive transformation of co-ordinates (symmetry). The passive sense, when the co-ordinate system is transformed to a different inertial frame, corresponds to a perspectival change in the manner just discussed. The active sense involves the physical system under consideration changing frames (given a ``boost"). The previous discussion shows that an active boost cannot be considered to be the system as an entity acquiring the velocity of the new frame because connected and unconnected lengths would behave differently (except in the case when the change in velocity was especially contrived but that restriction is not part of the concept of an active LT symmetry). Therefore an active LT ought to be pictured (if the process is physically pictured at all) as a transfer and re-assembly of the apparatus in the new frame. The connected lengths of the objects comprising the system then act as a guide for the placement of the unconnected objects.

\subsection{Constructive v principle approach}

Finally, leaving aside pedagogic considerations, we consider one of the arguments \cite{janssen} that STR should be presented as a principle or kinematic theory: the Lorentz invariance of the non-gravitational laws of physics can be traced back to the structure of Minkowski spacetime and so it is preferable and more logical to go directly from Minkowski spacetime to the results of STR (kinematically) than via the physical fields (dynamically). Put another way, if one insists on proceeding dynamically, the Lorentz invariance of all the physical fields is vital but the latter has its origin in the Minkowski spacetime so proceeding from the latter is preferable. \cite{laws} The argument that Lorentz invariance of the physical fields has its origin in the Minkowski spacetime is justified by the principle of common origin inference (COI). \cite{janssen} Put briefly, either one proposes that the Lorentz invariance of the three non-gravitational fields (the electromagnetic, weak and strong interactions) is a remarkable coincidence or one seeks a common origin and finds it in Minkowski spacetime. As Janssen mentions in passing \cite{janssen1}, an alternative is that the three non-gravitational fields might be different manifestations of a single field and it is that fact that is the COI for the Lorentz invariance of all the non-gravitational physical fields. There are quite strong arguments that the differences between the strong, weak and electromagnetic fields in the present era are due to symmetry breaking of a single field in an earlier era. \cite{wilczek3,wilczek} If the unification of the physical fields turns out to be correct, then the COI argument for the role of Minkowski spacetime in determining the Lorentz invariance of the physical fields involved in STR loses its force.  

\section{Conclusion}

Presenting STR from a purely kinematic point of view leaves students' understanding of the subject deficient and less able to deal confidently with non-text-book examples like the Dewan-Beran thought experiment. On the other hand, including the constructive or dynamical approach in presenting STR enriches student understanding of the subject. \cite{bell} The thought experiment involving the arrays of charges capable of acting as a measuring rod and a simple clock described here facilitates the presentation of STR from the constructive point of view.

On the more general question of whether STR is better viewed as a principle (kinematic) or constructive (dynamic) theory of physics, there does not seem to be any compelling reason which justifies the almost universal emphasis on STR as a principle theory. As for the question of whether one should take a dynamical or kinematical approach to the subject, it seems clear that the changes in a connected object (object in equilibrium under internal forces) when it is transferred between inertial frames involves dynamical effects. On the other hand, the different observations of a single connected object by different inertial observers is a perspectival, not a dynamical, effect.


\begin{thebibliography}{99}
\bibitem{reciprocity} In deriving Eq.~(7), it has been assumed that the speed of $S$ with respect to $\overline{S}$ is $-v$.  This follows from the method of clock synchronization we have assumed. It is less desirable to rely on the principle of reciprocity which requires attributing some properties to space and time which it would be best to avoid in a constructive approach to STR.
\bibitem{postulates} Sometimes the invariance of the speed of light for all inertial observers is made one of the two postulates in a presentation of STR but that is not the minimalist form of the postulates as originally proposed by Einstein.
\bibitem{gravfield} The acceleration could be achieved by immersing the apparatus in a uniform gravitational field for the required period of time (a gravitational field in which tidal effects are negligible over the dimensions of the apparatus).
\bibitem{xdirection} We consider only inertial observers whose relative velocity lies along the $\pm x$-axis.
\bibitem{laws} In the argument here, it is being tacitly assumed that if all the physical fields are Lorentz invariant, then all the laws of physics involving them must be Lorentz invariant.
\end{thebibliography}
\end{document}